\documentclass[fleqn,10pt]{wlscirep}
\usepackage[utf8]{inputenc}
\usepackage[T1]{fontenc}
\usepackage{bm}
\usepackage{graphicx}
\usepackage{multirow}

\title{Detection of Covid-19 From Chest X-ray Images Using Artificial Intelligence: An Early Review}

\author[1,*]{Muhammad Ilyas}
\author[2]{Hina Rehman}
\author[1,2]{Amine Nait-ali}
\affil[1,2]{UPEC, Science and Technology, France}
\affil[2]{UOM, Biotechnology, Malakand, France}

\affil[*]{e-mail: muhammad.ilyas@u-pec.fr}

\begin{abstract}
	In 2019, the entire world is facing a situation of health emergency due to a newly emerged coronavirus (COVID-19). Almost 196 countries are affected by covid-19, while USA, Italy, China, Spain, Iran, and France have the maximum active cases of COVID-19. The issues, medical and healthcare departments are facing in delay of detecting the COVID-19. Several artificial intelligence based system are designed for the automatic detection of COVID-19 using chest x-rays. In this article we will discuss the different approaches used for the detection of COVID-19 and the challenges we are facing. It is mandatory to develop an automatic detection system to prevent the transfer of the virus through contact. Several deep learning architecture are deployed for the detection of COVID-19 such as ResNet, Inception, Googlenet etc. All these approaches are detecting the subjects suffering with pneumonia while its hard to decide whether the pneumonia is caused by COVID-19 or due to any other bacterial or fungal attack.
\end{abstract}
\begin{document}

\flushbottom
\maketitle

\thispagestyle{empty}

\section{introduction}

In December 2019, COVID-19 pandemic emerged in wuhan china. It became a global health problem in a very short time because of it contact-transferred behaviours \cite{1,2}. COVID-19 is caused by a virus known as Severe Acute Respiratory Syndrome Coronavirus 2 or also SARS-Cov-2 \cite{3}. A large family of these virus are causing different kind diseases such as cold, Middle East Respiratory Syndrome and Severe Acute Respiratory Syndrome. The new addition to coronavirus family COVID-19 is discovered in 2019, which has never been detected in humans. COVID-19 is known as zoonotic disease because it is transferred from animal to human such as bat \cite{4}, similar to SARS-CoV virus which is contaminated through cat and MERS-CoV from  dromedary \cite{5}. It is presumed that respiratory transmission and physical contact is cause to spread COVID-19 rapidly. According to world economic forum , people with no symptoms or mild symptoms are the main reason of spreading the epidemic. one our of four subject shows no symptoms of COVID-19, although he/she is suffering with the disease \cite{6}. 

\begin{figure*}
    \centering
    \includegraphics[width=\textwidth]{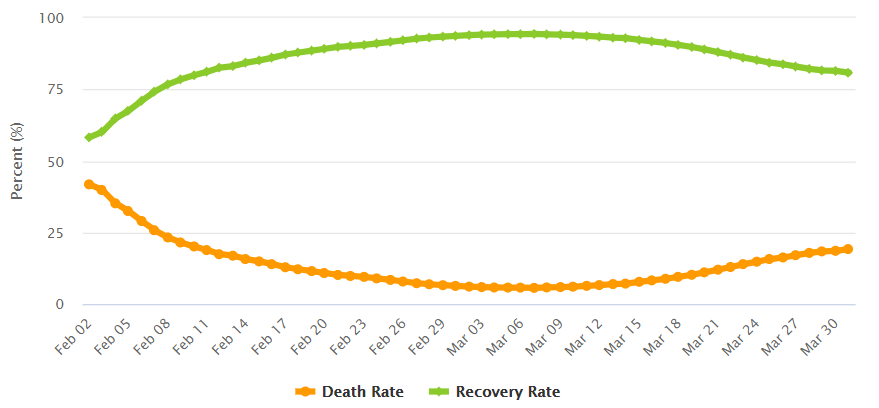}
    \caption{Death and recovery rate of COVID-19}
    \label{fig:death}
\end{figure*}

Nearly 82\% of the total infected subjects shows mild or no symptoms and the remaining are in critical conditions \cite{7}. Approximately 1,359,010 cases have been registered and 75,901 of them died and 293,454 were recovered till 7th April 2010. In the current status of infected subjects the statistics provides the probability that 95\% has the chances of recovery and 5\% is the mortality rate as shown in figure~\ref{fig:death} \cite{8}.

Dyspnea, fever and cough are symptoms of the infection while in more critical situation, this infection can lead to pneumonia, septic shock, SARS, organ failure and death \cite{4}. Comparative studies shows that the infection is widely spread in men as compare to women due to higher exposure and no deaths are reported in the children's in age range of 0-9 years \cite{7,8}. Respiratory problems spread faster in the subjects having pneumonia caused by COVID-19 as compared to healthy subjects \cite{9}.

\begin{figure*}
    \centering
    \includegraphics[width=\textwidth]{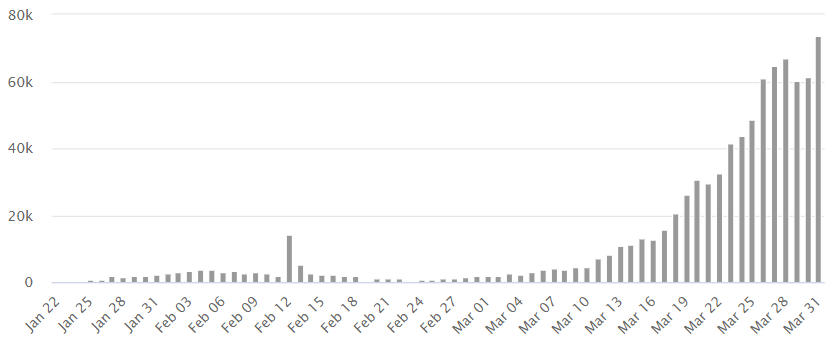}
    \caption{Daily cases of COVID-19}
    \label{fig:daily}
\end{figure*}

In the developing countries, the pandemic spread very rapidly even after taking the precautionary measure as shown in figure~\ref{fig:daily}. From march 19 till 31st march the number of infected subjects increased exponentially and the demand for intensive care units is increased in parallel \cite{8}.

The Chinese government publish the updated guidelines, COVID-19 could be diagnosed by gene sequencing through blood samples and Reverse Transcriptor Polymerase Chain Reaction (RT-PCR) is a specific indicator. The process gene sequencing using RT-PCR is time consuming and the subject should be hospitalized immediately.  

Thus considering the fact, the subjects tested positive with COVID-19 may have pneumonia and can easily be indicated by using automatic system for detection of COVID-19, and the concern department should consider immediate isolation and treatment for the subjects. A subject with critical stage, should suffered with a permanent lungs damage, if not died. According to world healthy organization, COVID-19 creates holes in the lungs similar to SARS which is not recoverable \cite{4} 

For detection of pneumonia the technique of computed tomography of the chest is also useful. Artificial Intelligence based system for automatic detection of COVID-19 can be helpful in monitoring, quantifying and distinguish a contact free subjective communication. A deep learning technique is also developed for extraction of graphical characteristics of COVID-19 from CT images to provide quick and precise diagnosis as compared to pathogenic testing and save the critical time \cite{10}. 

COVID-19 belongs to the same family of SARS-CoV and MERS-CoV, Scientific evidence supports the possibility to detect SARS-CoV and MERS-CoV using chest x-ray and CT images. Researchers have used the techniques of features extraction and data mining to identify the pneumonia caused by MERS-CoV and SARS-CoV \cite{11}. 

X-ray machines are normally used to scan the body for detection of fractured bones, tumors, pneumonia, and lungs infections while CT scanning is a Little advanced and more sophisticated system to examine different body part, tissues and organs more clearly. Using x-ray images is a bit cheap and easier way as compared to CT. While wrong detection may lead epidemic worst than expected \cite{12}. 

In this article, we will discuss the existing artificial intelligence based system for the detection of COVID-19 and the challenges these systems are facing. 

\section{Results}

Considering the challenges related to pandemic COVID-19, Artificial Intelligence (AI) can provide sophisticated solutions. The human knowledge, intelligence, and creativity along with the updated technology, its possible to beat the problems. 

The COVID-19 challenges are somehow exposing the drawback related to AI. The existing form of AI, in the form of machine learning and deep learning is trying to identify different pattern in the training databases. AI can provide sufficient results just in case having enough data for training and testing different systems with several approaches.

\begin{figure*}
    \centering
    \includegraphics[width=\textwidth]{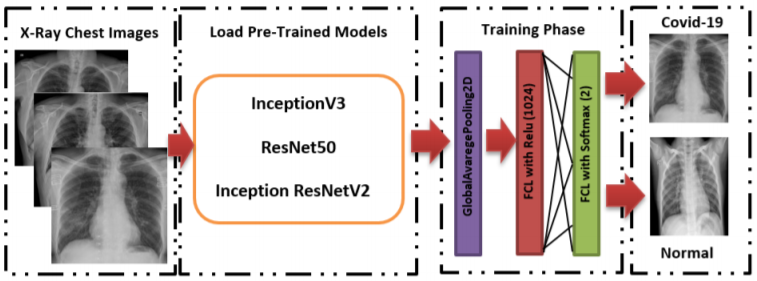}
    \caption{Schematic representation of pre-trained models for the prediction of COVID-19
patients and normal}
    \label{fig:model}
\end{figure*}
\begin{figure*}
    \centering
    \includegraphics[width=\textwidth]{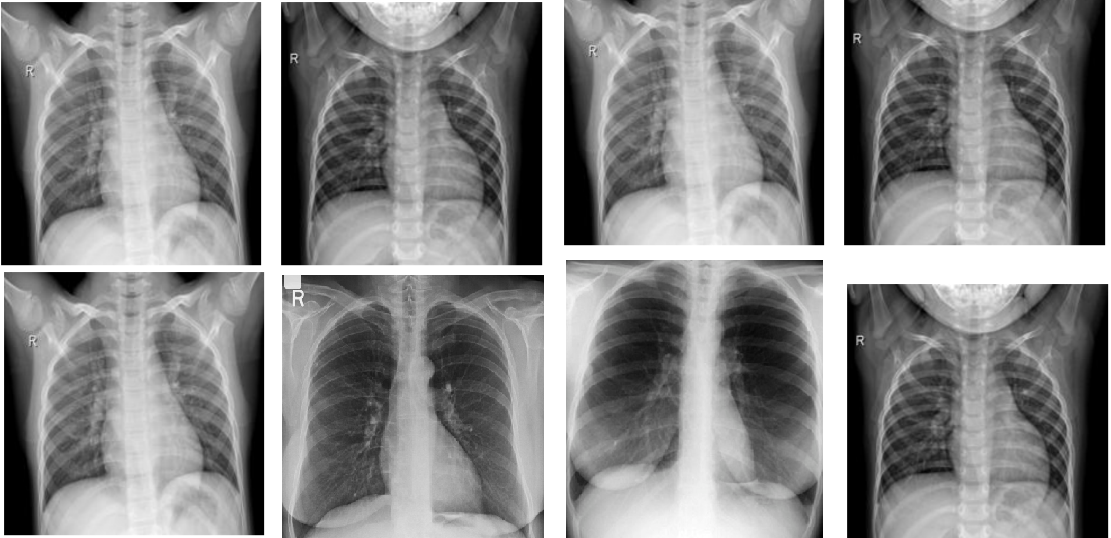}
    \caption{Representative chest X-ray images of normal}
    \label{fig:normal}
\end{figure*}
\begin{figure*}
    \centering
    \includegraphics[width=\textwidth]{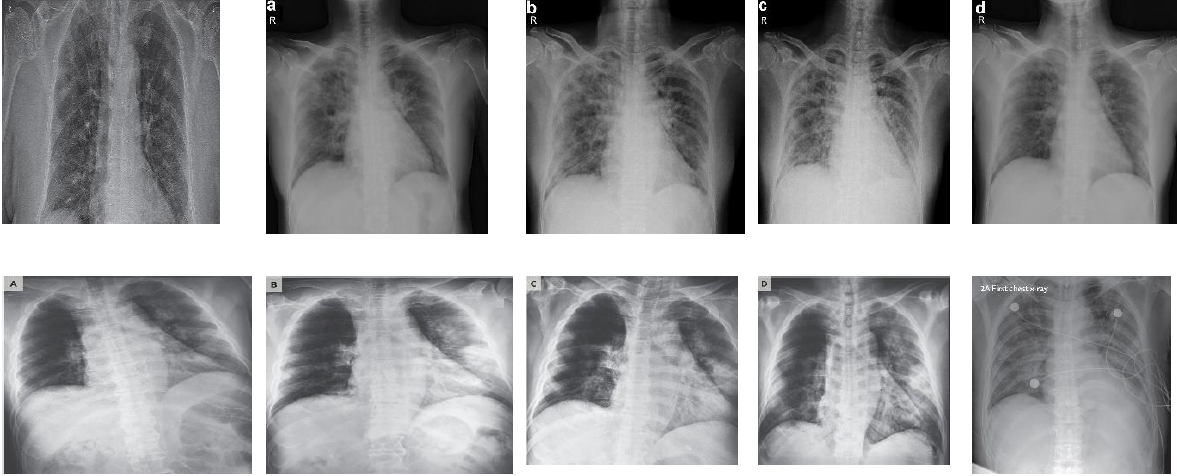}
    \caption{Representative chest X-ray images of COVID-19 patients}
    \label{fig:infected}
\end{figure*}

In table~\ref{table:comparison}, the comparison of different deep learning have evaluated on several databases \cite{18,19,20}.  
Ali et al. \cite{13} have used a dataset of nearly 100 subject \cite{18}, among them 50 x-ray images subjects were tested positive with COVID-19 and 50 x-ray images of normal subjects as shown in figure~\ref{fig:infected} and figure~\ref{fig:normal} respectively. They tried several deep learning approaches such as ResNet50, InceptionV3 and InceptionResNetV2 as shown in figure~\ref{fig:model}. Furthermore, ResNet50 model shows the highest accuracy of 98\% for the proposed dataset while InceptionV3 and Inception-ResNetV2 show 97\% and 87\% respectively. Prabira et al. \cite{14} used Support vector Machine (SVM) with different deep learning architecture (AlexNet, DenseNet201, GoogleNet, Inceptionv3, ResNet18, ResNet50, ResNet101 , VGG16, VGG19, XceptionNet and Inceptionresnetv2). They used a dataset of x-ray images consists of 50 images (25 subjects infected with COVID-19 and 25 subjects normal subjects). Among all the different architecture ResNet50 with 1000 feature vectors, gives the highest classification accuracy of 95\%. Jianpenget et al. \cite{15}, used deep anamoly detection model for detection of COVID-19. The database used for testing and traing consist of 100 x-ray images (70 subjects infected with COVID-19 while 30 normal subjects) \cite{18}. While they trained the system with 1431 chest x-ray having pneumonia to enhance the performance of the model to detect COVID-19. For the proposed model, the highest classification rate was achieved nearly 96\%. Ezz et al. \cite{16} used several deep learning architecture such as Visual Geometry Group Network (VGG), DenseNet, ResNetV2 etc. The dataset used is also limited to 50 images ( 25 infected subjects and 25 normal subjects). The deep neural network architecture used are able to identify the normalized intensities of the chest x-ray images for further classification.  
VGG19 and Dense Convolutions Networks shows similar accuracy for classification performance of 91\%. Ioannis et al. \cite{17} also presented similar approaches based on deep learning for detection of COVID-19. A database of nearly 728 x-ray images (224 positive subject with COVID-19 and 504 normal subject) \cite{20} are consider for training and testing purposes. They used the technique of transfer learning and achieved the classification accuracy of nearly 98\% using VGG19 deep learning architecture.


\begin{table*}[]
\centering
\begin{tabular}{llll}
\hline
Reference            & Model                               & Database size        & Accuracy \\ \hline
\multirow{3}{*}{\cite{13}}  & InceptionV3                         & \multirow{3}{*}{50 infected and 50 normal \cite{18}} & 96\%       \\ \cline{2-2} \cline{4-4} 
                     & ResNet50                            &                      & 98\%       \\ \cline{2-2} \cline{4-4} 
                     & InceptionV2                         &                      & 80\%       \\ \hline
\multirow{10}{*}{\cite{14}} & AlexNet                             & \multirow{10}{*}{25 infected 25 normal \cite{19}} & 93\%       \\ \cline{2-2} \cline{4-4} 
                     & DenseNet201                         &                      & 93\%       \\ \cline{2-2} \cline{4-4} 
                     & GoogleNet                           &                      & 91\%       \\ \cline{2-2} \cline{4-4} 
                     & InceptionV3                         &                      & 91\%       \\ \cline{2-2} \cline{4-4} 
                     & ResNet18                            &                      & 91\%       \\ \cline{2-2} \cline{4-4} 
                     & ResNet50                            &                      & 95\%       \\ \cline{2-2} \cline{4-4} 
                     & ResNet101                           &                      & 89\%       \\ \cline{2-2} \cline{4-4} 
                     & VGG16                               &                      & 92\%       \\ \cline{2-2} \cline{4-4} 
                     & XceptionNet                         &                      & 93\%       \\ \cline{2-2} \cline{4-4} 
                     & InceptionNetV2                      &                      & 93\%      \\ \hline
\cite{15}                   & ResNet Deep Anamoly detection model & 70 infected and 30 normal \cite{18}                  & 96\%       \\ \hline
\multirow{7}{*}{\cite{16}}  & VGG                                 & \multirow{7}{*}{25 infected and 25 normal \cite{19}}  & 90\%       \\ \cline{2-2} \cline{4-4} 
                     & DenseNet201                         &                      & 90\%       \\ \cline{2-2} \cline{4-4} 
                     & ResNetV2                            &                      & 70\%       \\ \cline{2-2} \cline{4-4} 
                     & Inception                           &                      & 50\%       \\ \cline{2-2} \cline{4-4} 
                     & InceptionResNetV2                   &                      & 80\%       \\ \cline{2-2} \cline{4-4} 
                     & Xception                            &                      & 80\%       \\ \cline{2-2} \cline{4-4} 
                     & MobileNetV2                         &                      & 60\%       \\ \hline
\multirow{5}{*}{\cite{17}}  & VGG19                               & \multirow{5}{*}{224 infected 504 normal \cite{20}} & 98\%       \\ \cline{2-2} \cline{4-4} 
                     & MobileNet                           &                      & 97\%       \\ \cline{2-2} \cline{4-4} 
                     & Inception                           &                      & 86\%       \\ \cline{2-2} \cline{4-4} 
                     & Xception                            &                      & 85\%       \\ \cline{2-2} \cline{4-4} 
                     & InceptionResNet                     &                      & 84\%       \\ \hline
\end{tabular}
\caption{Performance evaluation of different deep learning architectures}
\label{table:comparison}
\end{table*}
\section{Discussion}
In the existing AI techniques, all the approaches are detecting pneumonia caused by COVID-19 by using chest X-ray images database \cite{18,19,20}. Among the challenges AI is facing concerning the detection of pneumonia is "how the system realize that the pneumonia detection in the chest x-ray is caused by COVID-19", such that pneumonia can be caused by many other viruses, Bactria's and by fungus attack. 
In common practice, all the researcher agrees to problem while "the majority of deaths from COVID-19 are owing to pneumonia in the lungs of vulnerable patients,” said Dr. Tom Naunton Morgan, chief medical officer at behold.ai. “Pneumonia is a potentially life-threatening condition caused by several pathogens including, directly or indirectly, COVID-19 infection". The algorithms are trained to detect pneumonia in real time. 

The common type of pneumonia occurs outside of the health care centres or hospitals. The common reason of pneumonia is bacteria in most of the cases which is known as streptococcus. It occurs by itself and the symptoms are cold and flu and it can effect some portion of the lung, that condition is known as lobar pneumonia. There is also walking pneumonia which is not severe and do not requires any best rest. Fungal pneumonia is considered among the fatal infections that can cause death. It subjects with weak immune system may suffer alot with fungal pneumonia, and it is caused from absorbing organism. Soil and bird dropping is also a source of causes of the proposed fungal pneumonia. Viruses also causes pneumonia such SARS-Cov, MERS-CoV and newly emerged COVID-19. Some of the viruses that cause colds and the flu which leads to cause pneumonia. The most common causes of pneumonia are different kind of viruses which is absorb from different kind of animals. The reason to discuss these different kinds pneumonia and it causes is to justify the challenges to different different all them by using a strong AI system. 

AI can help the community in different areas such as early warnings and alerts, diagnosis and prognosis, tracking and prediction, treatments, and cures, data dashboards, and social control.

The blueDot \cite{22} a canadian AI model proved its mettle, and collected so much fame. According to the AI model, the user were informed about the outbreak on 31 December 2019 before World Health Organization (WHO) announced on 9th January 2020. Some researchers worked with BlueDot and published an article in Journal of Travel Medicine in January 2020 \cite{21}. They mentioned that the virus will be spread in 20 different cities in the world from Wuhan china by travelers. BlueDot was exaggerated as well underrated by several forums, although its a powerful AI tool. 
The second AI based model is Healthmap \cite{23}, which also gave an early alarm back on 31 December 2019 at Boston Children's Hospital in United States of America and a scientist also predicted a significant outbreak of the COVID-19 after 30 minutes. Although the AI model responded faster than human.

The precise detection with short computation time can save many lives. It can also control the spreading and generation of the disease through a well trained model. Scientists working with UN Global Pulse, reviewed several AI based application for the detection of COVID-19. The mentioned that the both x-ray and Computed Tomography (CT) scans can be used for the detection of COVID-19 based on AI models. Researchers also proposed the idea of using a mobile phone to scan CT images for the detection of COVID-19 \cite{24}.

For prediction and tracking purposes AI can also provide sufficient support such that in which manner that pandemic related to COVID-19 will be spread and in how much time. Following the previous pandemic in 2015, for Zika-virus a AI based system is developed to predict the spread of the disease \cite{25}. These existing model can be utilized for COVID-19 such that the system should be re-trained with data related to COVID-19. The algorithm trained for the prediction of seasonal flu can also be re-trained with new data from COVID-19 \cite{26}. For deep learning architecture we need a huge among of training data. In the state-of-the-art approaches, the databases used are very small. We cannot rely on the performance of a system which is train with 50-100 images with only one specific kind of pneumonia caused due to COVID-19. A big database of chest x-ray images, is required with different types of pneumonia in order to enhance the performance of the AI based system for the detection of COVID-19.

AI is contributing long before the COVID-19 outbreak in discovery of potentially emerging drugs. Considering the outbreak of COVID-19, several research lab indicated the need of AI to search the vaccine for the treatment. Scientist believe that the use of AI model can accelerate the process of searching cure and vaccine for COVID-19 such that The structure of the protein of COVID-19 was predicted by Google's DeepMind and it can provide useful information for the discovery of the vaccine. It also mention on the webpage of DeepMind that “we emphasize that these structure predictions have not been experimentally verified. We can’t be certain of the accuracy of the structures we are providing \cite{27}.”

The data dashboard is providing a sophisticated visualized information about the details of spreading of COVID-19. MIT technologies provided several lists to the AI based dashboard for tracking and forecasting the outbreak of COVID-19 such that  HealthMap \cite{28}, NextTrain \cite{29}, Upcode \cite{30} are included. All these dashboards provide a global view of COVID-19 outbreak in all the countries.

Lock-down and social distancing are considered the preliminary and basic precautionary measures by healthcare department. AI can be used to control the scanning of people in crowded or potentially effected area such that in the railway station in china infrared camera have been used to detect the human body temperature for the prediction of fever against COVID-19 \cite{31}.

AI has been, and can further be used, to manage the pandemic by scanning public spaces for people potentially infected, and by enforcing social distancing and lockdown measures. For example, as described in the South China Morning Post, “At airports and train stations across China, infrared cameras are used to scan crowds for high temperatures. They are sometimes used with a facial recognition system, which can pinpoint the individual with a high temperature and whether he or she is wearing a surgical mask”.

\section{Conclusion}

The proposed AI based approaches in the literature for detection of COVID-19 shows promising results such VGG19 with 98\% of accuracy, ResNET with 96\%, ResNet50 with 95\% of accuracy, and InceptionV3 with 96\%. The databases used in most of cases is about 50-100 x-ray images both infected subject with COVID-19 and normal subjects too. All the proposed approaches used a binary classification techniques, while pneumonia has several other causes too. 

To conclude the existing work, it is hard to fight COVID-19 because of its mysterious behaviour, and unknown biological origin. We can try the precautionary measures and lesson learned from other public health outbreaks such as SARS-CoV and MERS-CoV.  The awareness about wearing masks, social distancing, isolation, hygiene and quarantine can reduce the chances of spreading pandemic. Convalescent plasma is also considered as a potential therapy for COVID-19. 

The future challenges for AI related to the detection of COVID-19 is: training and testing different deep learning architectures with huge databases with all kind of chest x-ray images infected by different kinds of pneumonia. A multi-classification techniques is required with a data for more precise detection of COVID-19.



\begin{thebibliography}{references}

\bibitem{1} Roosa, K., Lee, Y., Luo, R., Kirpich, A., Rothenberg, R., Hyman, J. M., Yan, P., and
Chowell, G. Real-time forecasts of the COVID-19 epidemic in China from February 5th to
February 24th, 2020. Infectious Disease Modelling, 5 : 256-263, 2020.
\bibitem{2} Yan, L., Zhang, H. T., Xiao, Y., Wang, M., Guo, Y., Sun, C., Tang, X., Jing, L., Li, S.,
Zhang, M., Xiao, Y., Cao, H., Chen, Y., Ren, T., Jin, J., Wang, F., Xiao, Y., Huang, S., Tan,
X., Huang, N., Jiao, B., Zhang, Y., Luo, A., Cao, Z., Xu, H., and Yuan, Y. Prediction of
criticality in patients with severe Covid-19 infection using three clinical features: a machine
learning-based prognostic model with clinical data in Wuhan. medRxiv preprint doi:
https://doi.org/10.1101/2020.02.27.20028027, 1-18, 2020.

\bibitem{3} Stoecklin, S. B., Rolland, P., Silue, Y., Mailles, A., Campese, C., Simondon, A., Mechain,
M., Meurice, L., Nguyen, M., Bassi C., Yamani, E., Behillil, S., Ismael,S., Nguyen, D., Malvy,
D., Lescure, F. X., Georges, S., Lazarus, C., Tabaï, A., Stempfelet, M., Enouf, V., Coignard,
B., Levy-Bruhl, D. and Team, I. First cases of coronavirus disease 2019 (COVID-19) in France:
surveillance, investigations and control measures, January 2020. Eurosurveillance, 25(6)
:2000094, 2020

\bibitem{4} https://www.who.int/health-topics/coronavirus [Last visit 01.04.2020].

\bibitem{5} Huang, C., Wang, Y., Li, X., Ren, L., Zhao, J., Hu, Y., Zhang, L., Fan, G., Xu, J., Gu, X.,
Cheng, Z., Yu, T., Xia, J., Wei, Y., Wu, W., Xie, X., Yin, W., Li, H., Liu, M., Xiao, Y., Gao, 
H., Guo, L., Xie, J., Wang, G., Jiang, R., Gao, Z., Jin Q., Wang, J., and Cao, B. Clinical features
of patients infected with 2019 novel coronavirus in Wuhan, China. The Lancet, 395(10223):
497-506, 2020.

\bibitem{6} https://www.weforum.org/agenda/2020/03/people-with-mild-or-no-symptoms-could-be-spreading-covid-19/ [Last visited 01/04/2020]

\bibitem{7} https://www.nationalgeographic.com/science/2020/02/here-is-what-coronavirus-does-tothe-body/ [Last visited 01/04/2020]

\bibitem{8} https://www.worldometers.info/coronavirus/ [Last visited 01/04/2020]

\bibitem{9} Wang, Y., Hu, M., Li, Q., Zhang, X. P., Zhai, G., and Yao, N. Abnormal respiratory patterns
classifier may contribute to large-scale screening of people infected with COVID-19 in an
accurate and unobtrusive manner. arXiv preprint arXiv : 2002.05534, 1-6, 2020.

\bibitem{10} Wang, S., Kang, B., Ma, J., Zeng, X., Xiao, M., Guo, J., Cai, M., Yang, J., Li, Y., Meng,
X., and Xu, B. A deep learning algorithm using CT images to screen for Corona Virus Disease
(COVID-19). medRxiv preprint doi: https://doi.org/10.1101/2020.02.14.20023028, 1-26, 2020.

\bibitem{11} Xie, X., Li, X., Wan, S., and Gong, Y. Mining X-ray images of SARS patients. Data
Mining: Theory, Methodology, Techniques, and Applications, Williams, Graham J., Simoff,
Simeon J. (Eds.), pp. 282-294, ISBN: 3540325476 , Springer-Verlag, Berlin, Heidelberg, 2006.

\bibitem{12} https://biodifferences.com/difference-between-x-ray-and-ct-scan.html

\bibitem{13} Narin, A., Kaya, C.,  Pamuk, Z. (2020). Automatic Detection of Coronavirus Disease (COVID-19) Using X-ray Images and Deep Convolutional Neural Networks. arXiv preprint arXiv:2003.10849.

\bibitem{14} Sethy, P. K., Behera, S. K. (2020). Detection of coronavirus Disease (COVID-19) based on Deep Features.

\bibitem{15} Zhang, J., Xie, Y., Li, Y., Shen, C.,  Xia, Y. (2020). COVID-19 Screening on Chest X-ray Images Using Deep Learning based Anomaly Detection. arXiv preprint arXiv:2003.12338.

\bibitem{16} Hemdan, E. E. D., Shouman, M. A.,  Karar, M. E. (2020). COVIDX-Net: A Framework of Deep Learning Classifiers to Diagnose COVID-19 in X-Ray Images. arXiv preprint arXiv:2003.11055.

\bibitem{17} Apostolopoulos, I. D.,  Bessiana, T. (2020). Covid-19: Automatic detection from X-Ray images utilizing Transfer Learning with Convolutional Neural Networks. arXiv preprint arXiv:2003.11617.

\bibitem{18} https://github.com/ieee8023/covid-chestxray-dataset

\bibitem{19} Chest X-Ray Images (Pneumonia) https://www.kaggle.com/paultimothymooney/chest-xray-pneumonia

\bibitem{20} https://github.com/ieee8023/covid-chestxray-dataset

\bibitem{21} Bogoch, I. I., Watts, A., Thomas-Bachli, A., Huber, C., Kraemer, M. U.,  Khan, K. (2020). Pneumonia of Unknown Etiology in Wuhan, China: Potential for International Spread Via Commercial Air Travel. Journal of Travel Medicine.

\bibitem{22} https://bluedot.global/ [last visited 07/04/2020]

\bibitem{23} http://www.diseasedaily.org/about [last visited 07/04/2020]

\bibitem{24} Maghdid, H. S., Ghafoor, K. Z., Sadiq, A. S., Curran, K.,  Rabie, K. (2020). A novel ai-enabled framework to diagnose coronavirus covid 19 using smartphone embedded sensors: Design study. arXiv preprint arXiv:2003.07434.

\bibitem{25} Akhtar, M., Kraemer, M. U.,  Gardner, L. M. (2019). A dynamic neural network model for predicting risk of Zika in real time. BMC medicine, 17(1), 171.

\bibitem{26} https://www.technologyreview.com/s/615360/cdc-cmu-forecasts-coronavirus-spread/ [last visited 07/04/2020]

\bibitem{27} https://deepmind.com/research/open-source/computational-predictions-of-protein-structures-associated-with-COVID-19 [last visited 07/04/2020]

\bibitem{28} https://www.healthmap.org/covid-19/?mod=article\_inline [last visited 07/04/2020]

\bibitem{29} https://nextstrain.org/ncov [last visited 07/04/2020]

\bibitem{30} https://www.againstcovid19.com/singapore/dashboard [last visited 07/04/2020]

\bibitem{31} https://www.scmp.com/comment/opinion/article/3075553/time-coronavirus-chinas-investment-ai-paying-big-way [last visited 07/04/2020]


\end{thebibliography}

\end{document}